\documentclass[twocolumn,superscriptaddress,unsortedaddress,nofootinbib,amsmath,amssymb,
aps,floatfix]{revtex4-2}
\usepackage{graphicx,tikz} 
\usepackage[mode=buildmissing]{standalone} 
\usepackage{dcolumn}
\usepackage{bm}
\usepackage{physics}
\usepackage{xcolor}
\usepackage{siunitx}
\usepackage{soul}
\usepackage{comment}
\usepackage[bb=boondox]{mathalfa}
\usetikzlibrary{external}
\DeclareMathOperator{\sinc}{sinc}

\begin{document}

\title{Coulomb interaction-driven entanglement of electrons on helium}
\author{Niyaz~R.~Beysengulov}
\affiliation{Department of Physics and Astronomy, Michigan State University, East Lansing, MI 48824, USA}
\author{Øyvind S. Schøyen, Stian D. Bilek, Jonas B. Flaten, and Oskar Leinonen}
\affiliation{Department of Physics and Center for Computing in Science Education, University of Oslo, N-0316 Oslo, Norway}
\author{Morten Hjorth-Jensen}
\affiliation{Facility for Rare Isotope Beams and Department of Physics and Astronomy, Michigan State University, East Lansing, MI 48824, USA}
\affiliation{Department of Physics and Center for Computing in Science Education, University of Oslo, N-0316 Oslo, Norway}
\author{Johannes~Pollanen}
\affiliation{Department of Physics and Astronomy, Michigan State University, East Lansing, MI 48824, USA}
\author{Håkon Emil Kristiansen}
\affiliation{Department of Chemistry and Hylleraas Center for Quantum Molecular Sciences, University of Oslo, N-0316 Oslo, Norway}
\author{Zachary J. Stewart, Jared D. Weidman, and Angela K. Wilson}
\affiliation{Department of Chemistry, Michigan State University, East Lansing, MI 48824, USA}

\begin{abstract}
The generation and evolution of entanglement in many-body systems is an active area of research that spans multiple fields, from quantum information science to the simulation of quantum many-body systems encountered in condensed matter, subatomic physics, and quantum chemistry. Motivated by recent experiments exploring quantum information processing systems with electrons trapped above the surface of cryogenic noble gas substrates, we theoretically investigate the generation of \emph{motional} entanglement between two electrons via their unscreened Coulomb interaction. The model system consists of two electrons confined in separate electrostatic traps which establish microwave-frequency quantized states of their motion. We compute the motional energy spectra of the electrons, as well as their entanglement, by diagonalizing the model Hamiltonian with respect to a single-particle Hartree product basis. We also compare our results with the predictions of an effective Hamiltonian. The computational procedure outlined here can be employed for device design and guidance of experimental implementations. In particular, the theoretical tools developed here can be used for fine tuning and optimization of control parameters in future experiments with electrons trapped above the surface of superfluid helium or solid neon.
\end{abstract}


\date{\today}

\maketitle

\section{Introduction} 

Entanglement is the fundamental characteristic that distinguishes
interacting quantum many-body systems from their classical
counterparts. The study of entanglement in precisely engineered
quantum systems with countably many degrees of freedom is at the
forefront of modern physics, and it is a key resource in quantum
information science (QIS). This is particularly true in the
development of two-qubit logic for quantum computations, which has
been demonstrated in a wide variety of physical systems used in
present-day quantum computing, including superconducting
circuits~\cite{Steffen1423,Barends2014}, trapped
ions~\cite{Monroe1995,Schmidt-Kaler2003}, semiconductor quantum
dots~\cite{Li809,Petta2005,Reina2003,Reina2005}, color-center defects
in diamond~\cite{Dutt2007,Neumann2010,Bernien2013}, and neutral atoms
in optical lattices~\cite{Madjarov2020,
  evered2023highfidelity}. Investigating the generation and evolution
of entanglement in quantum many-body systems is also important for
quantum
simulations~\cite{feynman1982simulating,Trabesinger2012,Georgescu2014,Altman2021},
having the potential to advance the fundamental understanding of dense
nuclear matter or high-energy
physics~\cite{Carlson2018,Klco2020,Yamamoto2022,Illa2022,Bauer2022},
correlated electron systems~\cite{Smith2016,Hofstetter2018,Smith2019},
and quantum
chemistry~\cite{Peruzzo2014,Reiher2016,Kandala2017}. Quantum
simulators based on {\em natural} qubits such as
atoms~\cite{Greiner2002,Hart2015,Ebadi2021},
ions~\cite{Lanyon2011,Monroe2021} and photons~\cite{Aspuru-Guzik2012}
are particularly appealing since these systems are highly
programmable, controllable and
replicable~\cite{Alsing2023}. Additionally, in these systems the
coupling to decohering environmental degrees of freedom is minimal,
allowing for a tight feedback between experiments and theory.

Trapped electron systems represent a novel approach to investigating
the generation of entanglement, sharing many features with platforms
based on other natural qubit systems. Recent experimental efforts have
investigated the feasibility of trapped electron qubits using ion trap
techniques~\cite{Matthiesen2019,Yu2022}. In fact, the naturally
quantized motion of electrons trapped in vacuum above the surface of
superfluid helium was one of the earliest theoretical proposals for
building a large-scale analog quantum
computer~\cite{platzman1999quantum}. The surface of the superfluid
functions as a pristine substrate~\cite{Shirahama1995}, shielding the
electrons from deleterious sources of noise at the device layer
beneath helium. Since this initial proposal, a number of theoretical
ideas have been put forward to create both
charge~\cite{dykman2003qubits,Dahm2003,Schuster2010,Shi2014,Kawakami2023}
and
spin~\cite{Lyon2006,Schuster2010,Dykman2023,Kawakami2023,dykman2023spin}
qubits based on these trapped electrons. Additionally a wide variety
of experimental work, directed at realizing these electronic qubits,
has been performed to leverage advances in nano-fabrication techniques
for precision trapping and control of electrons on helium in confined
geometries~\cite{Marty1986,Ikegami2009,zhang2009,Rees2016b,Zou2022},
mesoscopic devices~\cite{Papageorgiou2004,Rees2011,Rees2012}, circuit
quantum electrodynamic
architectures~\cite{Yang2016,koolstra2019coupling}, and surface
acoustic wave devices~\cite{Byeon2021}. Single-electron trapping and
detection have been experimentally
achieved~\cite{Papageorgiou2004,Rousseau2007,koolstra2019coupling}, as
well as extremely high-fidelity electron transfer along gated arrays
fabricated using standard CMOS
processes~\cite{Bradbury2011}. Similarly, electrons trapped above the
surface of solidified noble gases offer an alternative trapped
electron qubit; electrons trapped in vacuum above the surface of solid
neon have recently been experimentally demonstrated as a novel natural
charge qubit~\cite{Zhou2022} with high coherence~\cite{Zhou2023}.
    
In aggregate, these technological advances have opened the door to
exploring the generation and evolution of entanglement in systems
based on trapped electrons. Here we present a model system for
investigating the entanglement between the microwave-frequency
motional states of two electrons trapped in vacuum above the surface
of a layer of superfluid helium. The electrons are confined laterally
by applying voltages to electrodes in a substrate beneath the
condensed helium layer. These voltages are tuned to set up
electrostatic traps on the helium surface to control the relative
position of the electrons and quantize their in-plane motional states
in the GHz-frequency range. We utilize the full configuration
interaction (CI for short in this work) method \cite{CramerFCI} for
distinguishable particles to compute the quantized motional
excitations of the system, as well as the entanglement between the
electrons generated by Coulomb interaction. These numerical studies
are in turn used to optimize the electrode voltages to maximise the
entanglement. We also present an effective theoretical model of the
two-electron system, as a useful tool to analyze the underlying
coupling mechanism between the electrons. Given the exact solution
provided by the CI calculations, we discuss the limitations of the
approximations of this effective model. Our work can be used to
provide feedback to future experimental realizations in which,
ultimately, control and readout of charged qubit states can be
achieved, via integration of microwave
resonators~\cite{Schuster2010,koolstra2019coupling,Zhou2022,Zhou2023}
using standard techniques based on circuit quantum electrodynamics
(cQED)~\cite{blais2020circuit}.
    
In section~\ref{sec:device} we present a schematic micro-device that
allows for controlled Coulomb-driven entanglement between two
electrons. We also describe a numerical procedure to find the optimal
parameters for this device to function as a two-qubit quantum
computer. Section~\ref{sec:results} contains our main results, with
detailed discussion of the system properties and comparison to an
effective model Hamiltonian. The final section contains conclusions,
perspectives, and outlook for future work. Additional details are
presented in various appendices.

\section{Device and Theory}\label{sec:device}
    
\begin{figure}
        \centering
        \includegraphics[width=\columnwidth]{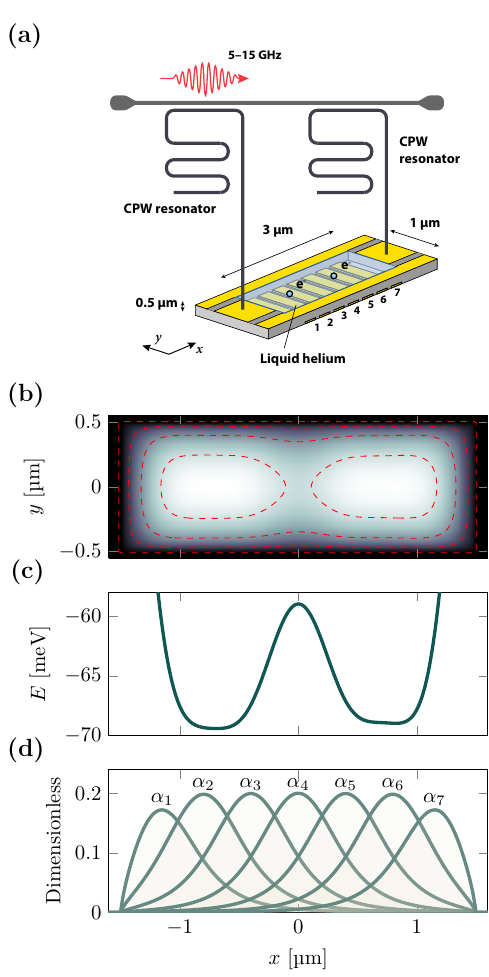}
        \caption{(a) Schematic micro-device, in which two electrons are trapped on the surface of a liquid helium basin in a double-well potential created by electrodes 1–7.
        Control and readout of the quantized motion is provided by two superconducting resonators, dispersively coupled to the in-plane motional states of the electrons.
        (b) Example configuration for the full two-dimensional electrostatic potential in the device, which realizes two separate wells.
        Brighter colors represent a stronger potential given in arbitrary
        units.
        (c) One-dimensional linecut of the potential in (b) along $y = 0$.
        (d) One-dimensional coupling constants from each individual electrode
        beneath the helium layer along $y = 0$.
        These coupling constants give rise to the total potential as given by Eq.~\eqref{eq:trap}.
        }
        \label{fig:anodes}
    \end{figure}

Electrons placed in vacuum above a layer of liquid helium are drawn
toward the liquid by an attractive force produced by positive image
charges in the dielectric liquid. However, the electrons are prevented
from entering the liquid by a large ($\sim$1~eV) Pauli barrier at the
liquid-vacuum interface~\cite{cole1969,shikin1971}. The balancing of
these two effects creates a ladder of Rydberg-like states for the
vertical motion of the electrons, and at low temperatures the
electrons are naturally initialized into the ground state of this
motion approximately \SI{11}{nm} above the helium
surface~\cite{Grimes76,Collin2002}. The electrons experience only a
weak interaction with their environment, which is mainly governed by
interactions with thermally excited ripplons (quantized capillary
waves on the helium surface) and phonons in the bulk of the liquid
\cite{wagner1973}. Based on these interactions, theory predicts long
coherence times of both the electron spin and motional degrees of
freedom~\cite{dykman2003qubits,lyon2006spin,Dykman2023}. The electron
in-plane motion can be further localized on length scales approaching
an electron separation of around $1$ µm through the integration of
micro-devices that provide lateral
confinement~\cite{rees2011point,rees2016structural,koolstra2019coupling}. Devices
of this type have been used to demonstrate single electron
trapping~\cite{Papageorgiou2005Counting,koolstra2019coupling,zhou2022single},
and to investigate the two-dimensional crystalline electronic phase
known as the Wigner
solid~\cite{rees2016structural,grimes1979evidence}, which arises from
the largely unscreened Coulomb repulsion between the electrons. As
explored in this work, this strong electron-electron interaction can
also in principle be utilized to couple the quantum motion of
electrons and create entanglement between electron charge qubits, in
analogy to a Cirac-Zoller entangling gate~\cite{CZ1995}.

\subsection{Device design}

A schematic micro-device for investigating the Coulomb-driven
entanglement of the in-plane motional states of two electrons on
helium is sketched in Fig.~\ref{fig:anodes}(a). Here we consider a $3\times 1$~µm$^2$
size microchannel structure with a depth of 0.5 µm,
filled with superfluid helium via capillary
action~\cite{marty1986stability}. Once the device is filled,
thermionic emission from a tungsten filament located above the helium
surface can be used to generate electrons, which are then naturally
trapped above the liquid surface. We note that trapping one or two
electrons also requires controlled loading and unloading of electrons
into the trap region from a larger reservoir area where electrons are
stored (not shown in Fig.~\ref{fig:anodes}(a)). This type of electron
manipulation is quite standard and has been experimentally
demonstrated in multiple devices, see for example
Refs.~\cite{bradbury2011efficient,koolstra2019coupling}. For the
purpose of the current theoretical study, we consider a simple array
of electrodes that allow for the investigation of entanglement between
two electrons, which we assume have already been loaded successfully
into the device. The rectangular device geometry and dimensions were
chosen to create an in-plane motional quantization axis along the
$x$-direction, with energy gaps in the frequency range of
$5$–$15$~GHz. These states are decoupled from motional states along
the $y$-direction at significantly higher frequency, which we will
ignore for the purposes of this one-dimensional study. Voltages
applied to seven $200$~nm-wide electrodes spaced by $200$~nm beneath
the helium layer provide the degrees of freedom needed to form an
electrostatic double-well potential for the two electrons as shown in
Figs.~\ref{fig:anodes}(b,c). The electrostatic potential in the trap
region is given
\begin{align}
        \varphi(x, y) = \sum_{i=1}^7 \alpha_i(x, y) V_i,
        \label{eq:trap}
\end{align}
where $\alpha_i = C_i/C_\Sigma$ is the relative contribution to the
potential defined by the capacitance between a region of space at
position $(x, y)$ on the helium surface and the corresponding
electrode. The total capacitance is $C_\Sigma = \sum_i C_i$, and $V_i$
is the voltage applied to the $i^\textrm{th}$ electrode, which can be
adjusted to create particular trapping potential configurations. We
note that the top electrodes at the helium surface are held at ground
potential. The coupling constants $\alpha_i (x, y)$ are calculated by
solving the Laplace equation for the electrostatic potential
numerically, using standard finite-element modeling techniques (see
Fig.~\ref{fig:anodes}(d)). The double-well trap is achieved by
applying a negative voltage to the central electrode (electrode 4 in
Fig.~\ref{fig:anodes}(a)) and positive voltages to the other
electrodes. Particular choices of applied voltages will be described
further in Section~\ref{sec:results}, where we also discuss how this
setup allows us to adjust the electron motional frequencies over a
broad range, enabling thereby the generation of entanglement between
the two electrons at certain conditions.

Coherent control and readout of the electron motional states in this
type of micro-device is based on coupling the electron motional states
to microwave frequency photons in superconducting resonators, see
Fig.~\ref{fig:anodes}(a), with a coupling $g_{\mathrm{RF}}/2 \pi =
\langle 1 | \mathbf{d} \cdot \mathbf{E}| 0 \rangle = e f_{\mathrm{RF}}
\partial \alpha_{\mathrm{RF}}/\partial x \sqrt{Z_{\mathrm{RF}}/m_e
  \omega_\text{e}}$. In this expression $\mathbf{d}$ is the dipole
moment of the oscillating electron along the $x$-axis, $\mathbf{E} =
\partial \alpha_{\mathrm{RF}}/\partial x \cdot V_{\mathrm{zpf}}
\hat{\mathbf{x}}$ is the electric field created by the resonator at
the position of the electron, $e$ and $m_e$ are the electron charge
and mass respectively, $\alpha_{\mathrm{RF}}$ is the coupling constant
for the resonator electrode, $V_{\mathrm{zpf}}$ is the voltage
amplitude of zero point fluctuations in the resonator,
$f_{\mathrm{RF}}$ and $Z_{\mathrm{RF}}$ are resonator frequency and
impedance, and $\omega_\text{e}$ is the electron motional frequency
along the $x$-axis. For typical values of $\partial
\alpha_{\mathrm{RF}}/\partial x = 0.5 \times 10^6$~m$^{-1}$,
$Z_{\mathrm{RF}} = 50$~$\Omega$, $f_{\mathrm{RF}} = 7$~GHz and
$\omega_\text{e}/2\pi = 5$~GHz, we find $g_{\mathrm{RF}}/2 \pi \simeq
12$~MHz.

At low temperatures, the decay of energy from the electrons-on-helium
system occurs due to its interaction with helium surface ripplons and
bulk phonons (see for example \cite{dykman2003qubits,Dykman2023}). The
total rate of decoherence due to these processes has been estimated to
be approximately $\Gamma/2\pi = 3 \times 10^4$
Hz~\cite{dykman2003qubits}, allowing the realization of the strong
coupling regime ($g_{\mathrm{RF}} \gg \Gamma$) between the microwave
photons and the electron motional states.

In this device, the two electrons are coupled individually to two
superconducting coplanar waveguide (CPW) $\lambda/4$-resonators, each
having a different resonant frequency. The crosstalk coupling between
an electron and the other electron's resonator is several times
smaller than the direct coupling to its own resonator, so we will
ignore this in our analysis. It should be noted that this classical
crosstalk can ultimately limit the fidelity of gate operations, but it
can be mitigated by applying appropriate compensation
tones~\cite{sheldon2016procedure}. In the dispersive regime of cQED,
in which $g_{\mathrm{RF}}/|\omega_\text{e} - \omega_\mathrm{RF}| \ll
1$, the frequency of the resonator is sensitive to the state of the
electronic motion, which can be detected by measuring the transmitted
microwave signals through the CPW feedline connected to the
resonators~\cite{Schuster2010,koolstra2019coupling,blais2020circuit}.

\subsection{Model Hamiltonian}

Our model Hamiltonian describes two electrons trapped in a double-well
potential set up by seven electrodes as given in Eq.~\eqref{eq:trap},
but we restrict our calculations along the $x$-direction only. The
interaction between the electrons is given by a Coulomb term which
gives rise to their correlated motion. The full Hamiltonian for the
system, in dimensionless units, is then given by

\begin{align}
        \hat{H}
        = \sum_{i = 1}^{2}\qty(
            -\frac{1}{2}\dv[2]{}{x_i}
            + v(x_i)
        )
        + u(x_1, x_2),
        \label{eq:hamiltonian}
\end{align}
where $v(x) = -e \varphi(x) /E_{\mathrm{d}}$ is the trap
potential. Here, $\varphi(x) = \varphi(x, 0)$ is the electrostatic
trap potential given in Eq.~\eqref{eq:trap}, and $E_{\mathrm{d}} =
\hbar^2/m_{\mathrm{e}} x_0^2$ is our energy unit ($\hbar$ is the
reduced Planck constant). The value $x_0 = 123$~nm is our length unit,
representing the characteristic inter-electron distance corresponding
to a typical electron density of $\simeq 2\times 10^9$~cm$^{-2}$ in
micro-devices~\cite{rees2016structural}. The soft Coulomb interaction
is given by
\begin{align}
        u(x_1, x_2)
        =
        \frac{\kappa}{\sqrt{\qty(x_1 - x_2)^2 + \epsilon^2}},
        \label{eq:soft-coulomb}
\end{align}
where $\kappa = e^2/(4 \pi \varepsilon_0 E_{\mathrm{d}}) = 2326$ gives
the strength of the Coulomb interaction ($\varepsilon_0$ is the vacuum
permittivity). We have introduced a shielding parameter $\epsilon =
10^{-2}$ to remove the singularity at $x_1 = x_2$~\cite{kvaal2007}. We
note that due to the small distance between the electrons and the
underlying electrodes, the Coulomb interaction will be reduced due to
screening effects. However in our analysis we consider an unscreened
Coulomb interaction, which sets an upper bound for the interaction
strength between the two electrons.

As long as the double-well potential $v(x)$ is sufficiently deep,
there will be no tunneling through the barrier between the wells for
the bound electron states.
   
This encourages us to split it into two separate potential wells.
Denoting the position of the barrier maximum by $x_b$, we can define

\begin{equation}
        \begin{gathered}
            v^L(x) = \begin{cases}
                v(x), & x < x_b, \\
                v(x_b), & x \geq x_b,
            \end{cases}
            \\
            v^R(x) = \begin{cases}
                v(x_b), & x < x_b, \\
                v(x), & x \geq x_b,
            \end{cases}
        \end{gathered}
        \label{eq:potential-splitting}
\end{equation}
with $L$ and $R$ labeling the left and the right wells
respectively. We can then express the total double-well potential as
the sum $v(x) = v^L(x) + v^R(x) - v(x_b)$. Since there is negligible
spatial overlap between single-electron states in different wells, we
can omit spin and focus on motional product states in which one
electron is localized in the left well while the other electron is
localized in the right well.
    
In essence, a sufficiently deep double-well trap allows us to treat
the electrons as distinguishable particles, labeled by their
position.\footnote{This claim was validated by also doing full
configuration interaction calculations with fermionic antisymmetry
between the electrons. All such calculations led to similar results as
the ones we present here with distinguishable electrons.} The one-body
Hamiltonian for each electron can then be written as
\begin{align}
        \hat{h}^A
        = -\frac{1}{2} \dv[2]{}{x}
        + v^A(x),
        \label{eq:one-body-hamiltonian}
\end{align}
with $A \in \qty{L, R}$, and the two-body Hamiltonian is given by Eq.~\eqref{eq:hamiltonian}.

Throughout our analysis we will vary the seven electrode voltages
$V_i$ to adjust the shape of the double-well potential $v(x)$, and
hence also the energy spectrum and frequencies of our system. We will
refer to each such choice as a \textit{well configuration}, and the
tuning between configurations is what allows us to realize various
quantum gates.

\subsection{State ansatz}

We solve the two-body problem described in the previous section by
exact diagonalization of the Hamiltonian in Eq.~\eqref{eq:hamiltonian}
with respect to a single-particle product basis. The two-body state
ansatz we use is
    \begin{align}
        \ket*{\Phi_n}
        &= \sum_{i = 0}^{N^L} \sum_{j = 0}^{N^R} C_{ij, n}
        \ket*{\phi^{L}_i \phi^{R}_j}.
        \label{eq:wave-function-ansatz}
    \end{align}
Here, $n$ is the index of each two-body energy eigenstate, and
$\ket*{\phi^L_i \phi^R_j} = \ket*{\phi^L_i} \otimes \ket*{\phi^R_j}$
are two-body product states built from two single-particle basis sets
$\qty{\ket*{\phi^A_i} \mid i = 0, \dots, N^A}$ (with $A \in \qty{L,
  R}$). The above ansatz is analogous to the ansatz of full
configuration-interaction (CI) theory, but since our electrons are
effectively distinguishable we use separable product states instead of
anti-symmetrized Slater determinants in our
expansion~\cite{CramerFCI}.

The quality of the ansatz in Eq.~\eqref{eq:wave-function-ansatz}
depends on the choice of single-particle basis states
$\ket*{\phi^A_i}$.  Even though we consider only two particles, a
large single-particle basis will quickly make the exact
diagonalization procedure prohibitively time consuming.  This limits
us to consider small single-electron basis sets, whose product states
span the state space of our two-electron system to a good
approximation.  One option is to consider the eigenstates of the
individual one-body Hamiltonians $\hat{h}^A$ defined in
Eq.~\eqref{eq:one-body-hamiltonian}. However, this approach neglects
all information about interactions and, as a consequence, still
demands a significant number of basis states to accurately capture the
physics.  A more effective approach is to employ the Hartree method
(analogous to the Hartree-Fock method, but for distinguishable
particles), which incorporates the one-body Hamiltonian with a
mean-field contribution from the Coulomb interaction.  This method has
the advantage of producing single-particle basis sets that can be
truncated to only a few states while still capturing the interaction
physics of the entangled two-body states within our system.
  
The construction of the Hartree basis sets $\ket*{\phi^A_i}$ and
derivation of the Hartree method are presented in
Appendices~\ref{app:basis-set}~and~\ref{app:Hartree-method}.  With the
single-particle basis sets established, the coefficients $C_{ij, n}$
in Eq.~\eqref{eq:wave-function-ansatz} can be calculated to find the
full two-body energy eigenstates for each well configuration. This is
done through a diagonalization procedure, which is explained in detail
in Appendix~\ref{app:fci}.

We should add, as discussed in more detail in
appendices~\ref{app:basis-set}~and~\ref{app:fci}, that we also have
performed full configuration interaction calculations with an
anti-symmetrized wave function basis for the two-electron system. For
the system we are investigating, the Hartree ansatz with
distinguishable particles gives an excellent approximation to the
anti-symmetrized full configuration interaction calculations.

\subsection{Entanglement}\label{sec:entanglement}

It is natural to consider the system at hand as bipartite, composed of
the two electrons as, ideally, individual subsystems. Such a
bipartition comes with the notion of entanglement --- the inability to
discern the exact state of each subsystem, even though the state of
the full system is known. We aim to find certain well configurations
for which a subset of the energy eigenstates are entangled, in order
to enable the set up of two-qubit gates, see for example the
discussions in
Refs.~\cite{Reina2005,strauch2003,dicarlo2009demonstration}.
   
A common entanglement measure for bipartite systems is the von Neumann
entropy of a quantum state, defined as
    \begin{equation}
        S = -\tr\!\qty(\hat{\rho}\log_2(\hat{\rho})),
    \end{equation}
where $\hat{\rho}$ is the reduced density operator of either
subsystem. We use this measure to quantify entanglement and will refer
to it simply as the entropy. (See
appendix~\ref{app:von-neumann-entropy} for calculational details.) In
what follows we denote the entanglement entropy of each energy
eigenstate $\ket*{\Phi_n}$ by $S_n$.

The two-body state of the full system can be expanded in any product
state basis from the subsystems, such as in
\eqref{eq:wave-function-ansatz}. While the Hartree basis discussed
above provides a succinct picture of the \textit{interaction} between
subsystems, another basis offering a clear picture of the
\textit{entanglement} is the Schmidt basis, found by doing a
singular-value decomposition of the coefficient matrix $C_{ij,n}$ as
outlined in appendix~\ref{app:von-neumann-entropy}. In the Schmidt
decomposition of a two-body state, each term involves a product of
unique, orthogonal Schmidt states. It follows that the Schmidt states
are eigenstates of the reduced density operators of each
subsystem. Then, the mixed state of each subsystem can be interpreted
as a statistical ensemble of its Schmidt states, and the Schmidt
coefficients (the singular values), when squared, give the occupation
number of each Schmidt state.  Our calculations indicate that the
Hartree basis actually serves as an approximate common Schmidt basis
for all the two-body energy eigenstates of our system. In addition to
the von Neumann entropies $S_n$, we therefore map the two-body
coefficients $C_{ij,n}$ to provide a clear overview of which products
of single-electron states that are involved in each entangled two-body
state. For simplicity, we will denote the Hartree product states as
    \begin{equation}
        \ket*{\phi^L_i \phi^R_j} = \ket{ij}, \label{eq:prod-state-notation}
    \end{equation}
but note that these product states are not to be directly interpreted
as computational basis states for quantum computing. We can not do any
measurements to collapse the two-electron system into any of these
separable states, so they should be interpreted only as an ideal
single-particle product basis for describing the two-body states of
our system. The states that \textit{should} be interpreted as
computational basis states, are four specific energy eigenstates of
configuration I, as defined in section~\ref{sec:gate-operation} below.

\subsection{Gate operation}\label{sec:gate-operation}

We target three specific well configurations that are ideal for
operation of one-qubit rotations as well as two-qubit
$\sqrt{i\mathrm{SWAP}}$ and CZ gates~\cite{nielsen2010quantum,
  strauch2003, triple-avoided-crossing}. Each configuration is defined
through specific entanglement entropies of the two-body energy
eigenstates.

Configuration I, see also the discussions in the next subsection and
Fig.~\ref{fig:target-configs}, corresponds to the case in which each
electron has a distinct transition frequency between its ground and
first-excited state. The correlations between the two electrons are
then minimal, and the state of the electrons can be controlled and
read out independently via their associated resonators.  We focus on
cases in which the frequency of the left qubit is larger than that of
the right qubit, and within the resonator working range of 5--15
GHz. Then the two-body energy eigenstates $\ket*{\Phi_0}$,
$\ket*{\Phi_1}$, $\ket*{\Phi_2}$ and $\ket*{\Phi_4}$ have maximum
overlap with the Hartree product states $\ket*{00}$, $\ket*{01}$,
$\ket*{10}$ and $\ket*{11}$, respectively, and we interpret these
eigenstates as computational basis states. Due to the minimized
correlation, the entanglement entropy is zero for all energy
eigenstates of this configuration.
   
Configuration II is designed to realize the two-qubit
$\sqrt{i\mathrm{SWAP}}$ gate. It can be achieved by an avoided
crossing of the first and second excited eigenstates, so that they are
given by
    \begin{equation}
        \begin{aligned}
            \label{eq:target-states-II}
            \ket*{\Phi_1} &= \qty(\ket*{10} - \ket*{01})/\sqrt{2},\\
            \ket*{\Phi_2} &= \qty(\ket*{10} + \ket*{01})/\sqrt{2}.
        \end{aligned}
    \end{equation}
All other energy eigenstates must remain product states to ensure that
only $\ket*{10}$ and $\ket*{01}$ are coupled. The entropy is then $1$
for the two states $\ket*{\Phi_1}$ and $\ket*{\Phi_2}$ and zero for
the rest. For a further discussion of avoided level crossings in
coupled quantum dot systems, see for example Ref.~\cite{Reina2005}.

The presence of higher energy levels gives rise to a different type of
correlation between the two electrons in our system. We are
particularly interested in a specific type of interaction that enable
the realization of a controlled-phase CZ gate
\cite{blais2020circuit,triple-avoided-crossing}. Configuration III
realizes the conditions to implement this type of two-qubit gate,
which involves a ``triple'' avoided crossing between the third, fourth
and fifth excited eigenstates:
    \begin{equation}
        \begin{aligned}
            \label{eq:target-states-III}
            \ket*{\Phi_3} &= \qty(\ket*{20} - \ket*{02} - \sqrt{2} \ket*{11})/2,\\
            \ket*{\Phi_4} &= \qty(\ket*{20} + \ket*{02})/\sqrt{2},\\
            \ket*{\Phi_5} &= \qty(\ket*{20} - \ket*{02} + \sqrt{2} \ket*{11})/2.
        \end{aligned}
    \end{equation}
The entropies of these states are $1.5$, $1$ and $1.5$,
respectively. In this configuration, too, the remaining energy
eigenstates must stay as close to their non-interacting counterparts
as possible, with entropy close to zero. To quantify the strength of
this two-qubit interaction, we define the $ZZ$ coupling strength
\cite{dicarlo2009demonstration,zz-suppression,zz-suppression2}, see
also discussions below, as
    \begin{equation}
        \label{zz-coupling}
        \zeta = E_4 - E_2 - E_1 + E_0.
    \end{equation}
This quantity measures the shift in transition frequency of one
electron when the other electron is excited. The $ZZ$-coupling plays
an important role in our analysis since it conveys information about
the coupling to higher excited states.
    
In configurations I and II, $\zeta$ should be as small as possible to
minimize phase errors when driving gates.  However, in configuration
III the action of the shift can be used to alter the phase of the
computational basis state $\ket*{\Phi_4}$, which generates the CZ
gate~\cite{strauch2003,zz-suppression,triple-avoided-crossing}.

Tuning the electrode voltages diabatically between configuration I and
configurations II or III, the above-mentioned two-body quantum gates
are realizable~\cite{strauch2003,triple-avoided-crossing}. We proceed
to demonstrate that the necessary well configurations and resulting
electron entanglement are achievable through targeted numerical
optimization. The remainder of this work is focused on the resulting
configurations, and we will leave actual simulations of time-dependent
gate operation to future work.

\subsection{Configurational search}\label{sec:config-search}

The three desirable well configurations defined in the previous
section can be targeted through numerical optimization methods, with
the seven electrode voltages as the variational parameters. To achieve
the avoided crossings described above, we can use the fact that the
Hartree product basis incorporates much of the Coulomb interaction
between the electrons, so that the residual Coulomb interaction term
is small. This means that the energy spectrum of the full interacting
system should be close to the spectrum of Hartree product
states\footnote{In our discussions we will refer to the Hartree states
as our idealization of the non-interacting system. This is not
entirely correct since the Hartree states do include correlations from
the Coulomb interaction.} $\ket*{ij}$, with transition energies given
by the sum of the corresponding Hartree transition energies
$\epsilon^L_i-\epsilon^L_0+\epsilon^R_j-\epsilon^R_0$ (where
$\epsilon^A_i$ are the eigenvalues of the Hartree states as defined in
appendix~\ref{app:Hartree-method}). This approximation matches the
interacting spectrum well except at the avoided crossings, where the
interaction turns what would have been an energy crossing in the
non-interacting case into an avoided crossing in the interacting
case. In other words, we can look at the Hartree energies of the
system and target degenerate Hartree energies to find avoided
crossings.
    
We also target qubit anharmonicities of equal magnitude but opposite
sign throughout all three configurations. This was shown to suppress
the unwanted $ZZ$-coupling defined in \eqref{zz-coupling} for
superconducting
qubits~\cite{triple-avoided-crossing,zz-suppression,zz-suppression2},
and so we investigate if the same principle is applicable to our
charge qubits. The anharmonicity of each qubit can again be defined
through the Hartree energies, which serve as a non-interacting,
single-particle guiding picture throughout this section.
\begin{figure*}
        \centering
        \includegraphics[width=2.0\columnwidth]{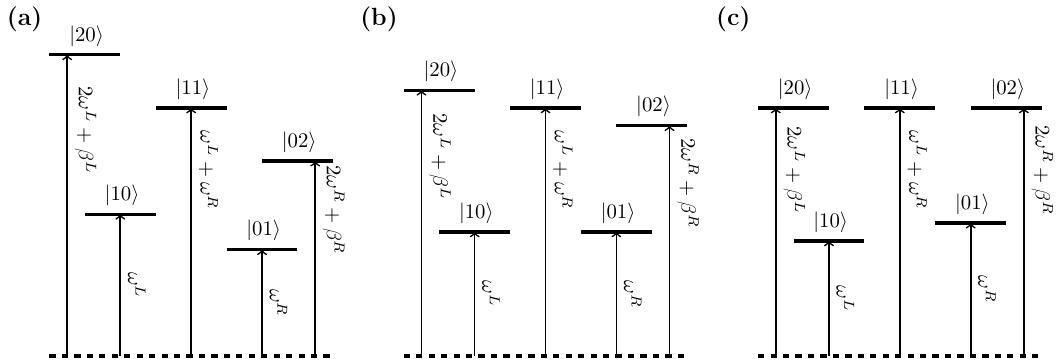}
        \caption{%
            (a) Transition frequencies in the non-interacting picture for configuration I. This configuration corresponds to a detuned system in which all transition energies are distinguishable, and $\Delta_\mathrm{I}= \omega^L - \omega^R > 0$, with $\beta^L = -\beta^R = \Delta_\mathrm{I} / 2$.
            (b) Transition frequencies in configuration II. Here the detuning is zero, $\Delta_\mathrm{II} =  0$, and the two states $\ket*{10}$ and $\ket*{01}$ are degenerate in the absence of interactions. Distinct anharmonicities kept at $\beta^L = -\beta^R = \Delta_\mathrm{I} / 2$ separate the higher states from one another.
            (c) Configuration III is realized when the three states $\ket*{20}$, $\ket*{11}$ and $\ket*{02}$ share the same transition frequency from the ground state. 
            This occurs when $\beta^L = -\beta^R = -\Delta_\mathrm{III}$. The detuning has opposite sign of that in configuration I, $\Delta_\mathrm{III} = -\Delta_\mathrm{I}/2$.
        }
        \label{fig:target-configs}
\end{figure*}
    
Figure~\ref{fig:target-configs} illustrates the non-interacting energy
spectra of the three target configurations.  The transition frequency
from $\ket*{0}$ to $\ket*{1}$ for subsystem $A \in \qty{L, R}$ is
denoted by $\epsilon^A_1-\epsilon^A_0 = \omega^A$ (with $\hbar =
1$). In order to selectively address the ground and first excited
energy eigenstates while avoiding population of higher states, the
electrostatic potential is intentionally designed to be anharmonic.
We define the \emph{anharmonicity} to be the difference in the
excitation energy between $\ket*{0} \to \ket*{1}$ and $\ket*{1} \to
\ket*{2}$. Consequently, the transition frequency for $\ket*{0} \to
\ket*{2}$ is given by $ \epsilon^A_2 - \epsilon^A_0 = 2\omega^A +
\beta^A$, where $\beta^A$ is the anharmonicity.
    
The energy of the non-interacting Hartree product state $\ket*{ij}$ is
given by $\epsilon_{ij} = \epsilon^L_i + \epsilon^R_j$.  We refer to
the difference in energy between the states $\ket*{10}$ and
$\ket*{01}$ as the qubit \emph{detuning}, and denote it by $\Delta
\equiv \omega^L - \omega^R$.  Using the detuning and the anharmonicity
we can express the transition frequencies for $\ket*{11} \to
\ket*{20}$ and $\ket*{02} \to \ket*{11}$ by $\epsilon_{20} -
\epsilon_{11} = \Delta + \beta^L$ and $\epsilon_{11} - \epsilon_{02} =
\Delta - \beta^R$.

Figure~\ref{fig:target-configs}(a) illustrates the non-interacting
energy spectrum for configuration I. In this configuration all
transition frequencies are distinct, and we have chosen a detuning of
$\Delta_{\mathrm{I}} = \omega^L - \omega^R > 0$ so that the electron
in the left well has higher transition frequencies than the electron
in the right well.  Furthermore, we have set $\beta^L = -\beta^R =
\Delta_{\mathrm{I}} / 2$ such that $\Delta + \beta^L = \Delta -
\beta^R > \Delta$, i.e., the energy gaps between $\ket*{20}$ and
$\ket*{11}$, and $\ket*{11}$ and $\ket*{02}$ are equally large, and
greater than the detuning.

Figure~\ref{fig:target-configs}(b) shows the target non-interacting
energy spectrum for configuration II. In this configuration the
single-particle basis states $\ket*{01}$ and $\ket*{10}$ are
degenerate, while the higher states $\ket*{20}$, $\ket*{11}$ and
$\ket*{02}$ are kept separate from each other.  This implies that
$\Delta_{\mathrm{II}} = \omega^L - \omega^R = 0$, and we have
maintained the anharmonicities at the same values as in configuration
I, i.e., $\beta^{L} = -\beta^{R} = \Delta_{\mathrm{I}} / 2$.
    
Finally, Fig.~\ref{fig:target-configs}(c) shows the target
non-interacting energy spectrum for configuration III. In this
configuration the higher states $\ket*{20}$, $\ket*{11}$ and
$\ket*{02}$ are degenerate, while $\ket*{10}$ and $\ket*{01}$ are
distinct. To realize this configuration we require $\Delta + \beta^L =
\Delta - \beta^R = 0$, and with $\beta^L = -\beta^R$ we find $\beta^L
= -\Delta$.  Keeping the anharmonicities of the two wells the same as
in configuration I and II, i.e., $\beta^L = -\beta^R =
\Delta_{\mathrm{I}} / 2$, leads to $\Delta_{\mathrm{III}} =
-\Delta_{\mathrm{I}} / 2$.

The residual Coulomb interaction between the electrons splits the
degeneracy in energy levels and leads to avoided level crossings,
entanglement between the two electrons, and hence the possibility of
driving two-qubit gates.  The anharmonicities being non-zero, with
equal magnitude and opposite sign also ensure that the avoided
crossing between the first and second excited states and the triple
avoided crossing between the higher states are
separated~\cite{triple-avoided-crossing, zz-suppression}.

We note, as indicated in Figs.~\ref{fig:target-configs}(a,c), that the
detunings in configuration I and configuration III have opposite
signs. This is not incidental, but has a deliberate purpose; it allows
for the realization of configuration II somewhere in the transitional
region between configuration I and III, as long as the anharmonicities
have a magnitude greater than zero along the same path.  This happens
because the detuning has to change sign in order to go from
configuration I to configuration III, leading to the characteristic
level crossing of configuration II when the detuning is zero.  Hence,
our task simplifies to locating configurations I and III with equal
anharmonicities by tuning the electrode voltages.  We can then define
a parametrization that interpolates between these two configurations,
and as long as the anharmonicities do not go to zero, we are
guaranteed to get a configuration II somewhere along the
parametrization path.

\section{Results and Discussion}\label{sec:results} 

We start this section by summarizing the numerical optimization
procedure that was used to locate the above-defined configurations I
and III in the parameter space of seven electrode voltages. We then
define a parametrization of the voltages and identify the location of
configuration II. Thereafter we discuss the properties of each
configuration in more detail. Finally, we make an attempt at
interpreting our results in terms of a phenomenological model.

\subsection{Configurational results}

To find the electrode voltages corresponding to configurations I and
III, we express the search as an optimization problem by defining cost
functions whose minima align with the desired properties for each
configuration, as described in section \ref{sec:config-search}.  Each
cost function was minimized by evaluating its gradient with respect to
the voltages. The optimization of the cost functions was done using
standard gradient descent methods with the so-called ADAM
algorithm~\cite{adamoptimizer} for the gradient updates.  As is common
in the optimization of multi-parameter functions, we found that our
cost functions often exhibit several local minima, a feature which
makes our solution dependent on the initial guess for the
voltages. Because of this, our approach involved manually adjusting
the voltages to obtain an initial well configuration resembling a
double-well trap with features close to the desired properties, and
then running the optimization search. Appendix~\ref{app:opt-well-conf}
provides an in-depth discussion of the full optimization process,
including specific expressions for the cost functions.

For configuration I, this procedure successfully achieves distinct
transition frequencies of each well, within the resonator working
range of 5--15 GHz. We also target anharmonicities with equal
magnitude and opposite signs to suppress $ZZ$ crosstalk in higher
energy states as discussed in
section~\ref{sec:config-search}. However, an arbitrary choice of
transition frequencies and anharmonicities does not necessarily result
in an appropriate well configuration. By performing the optimization
process for a range of possible candidates, we ended up targeting the
specific transition frequency of $\omega^L/2\pi = \SI{11}{\giga
  \hertz}$ between the two lowest energy levels in the left well, and
a transition frequency of $\omega^R/2\pi = \SI{9}{\giga \hertz}$ in
the right well. This corresponds to a detuning of $\Delta_\mathrm{I} /
2\pi = (\omega^L - \omega^R)/2\pi = \SI{2}{\giga\hertz}$.  At the same
time, anharmonicities of $\beta^L/2\pi = -\beta^R/2\pi =
(\Delta_{\mathrm{I}} / 2)/2\pi = \SI{1}{\giga\hertz}$ were
targeted. Optimization of the cost function based on these target
values (Eq.~\eqref{eq:LossConfigurationI} in
Appendix~\ref{app:opt-well-conf}) yields properties that are very
close to the desired ones. The two-body energies of the resulting
configuration are $E_1/2\pi = \SI{8.99}{\giga\hertz}$ and $E_2/2\pi =
\SI{11.01}{\giga\hertz}$ relative to the ground state, and the
anharmonicities are equal to the targeted values of
\SI{\pm1}{\giga\hertz} to three decimal places.

For configuration III, we achieve a triple degeneracy point between
the computational basis state $\ket{11}$ and the states $\ket{20}$ and
$\ket{02}$.  Here we construct a cost function targeting the entropies
of the energy eigenstates $\ket{\Phi_3}$, $\ket{\Phi_4}$ and
$\ket{\Phi_5}$ to be $1.5$, $1.0$ and $1.5$ respectively, while
keeping the entropies of all other eigenstates small.  In addition, we
target the detuning $\Delta_\mathrm{III}/2\pi$ to be $\SI{-1}{\giga
  \hertz}$ and the same anharmonicities as for configuration I,
$\beta^L/2\pi = -\beta^R/2\pi = \SI{1}{\giga\hertz}$. As discussed
earlier, this guarantees the presence of configuration~II for a
certain set of voltages in the transition from configuration~I to
configuration~III. We use the set of voltages obtained for
configuration I as an initial guess for the optimization of this cost
function (Eq.~\eqref{eq:LossConfigurationIII} in
Appendix~\ref{app:opt-well-conf}). This optimization results in
two-body energies $E_3/2\pi = \SI{27.24}{\giga\hertz}$, $E_4/2\pi =
\SI{27.42}{\giga\hertz}$ and $E_5/2\pi = \SI{27.61}{\giga\hertz}$
relative to the ground state, with entropies of 1.50, 1.00 and 1.49,
respectively.

To visualize properties of the configurations and the tuning between
them, we express the seven electrode voltages with one configuration
parameter $\lambda$, through a linear parametrization
    \begin{align}
        \vb*{V}(\lambda)
        = (1-\lambda)\vb*{V}_\mathrm{I} + \lambda\vb*{V}_\mathrm{III}.
        \label{eq:well-parameterization}
    \end{align}
Here, $\vb*{V}_\mathrm{I}$ and $\vb*{V}_\mathrm{III}$ are vectors with
the optimized voltages for configurations I and III. By construction,
configuration I then corresponds to $\lambda = 0$ while configuration
III corresponds to $\lambda = 1$. Explicit values of the voltages for
each optimized configuration are provided in table~\ref{tab:voltages}
in Appendix~\ref{app:voltages}.
\begin{figure}
    \centering
    \includegraphics[width=\columnwidth]{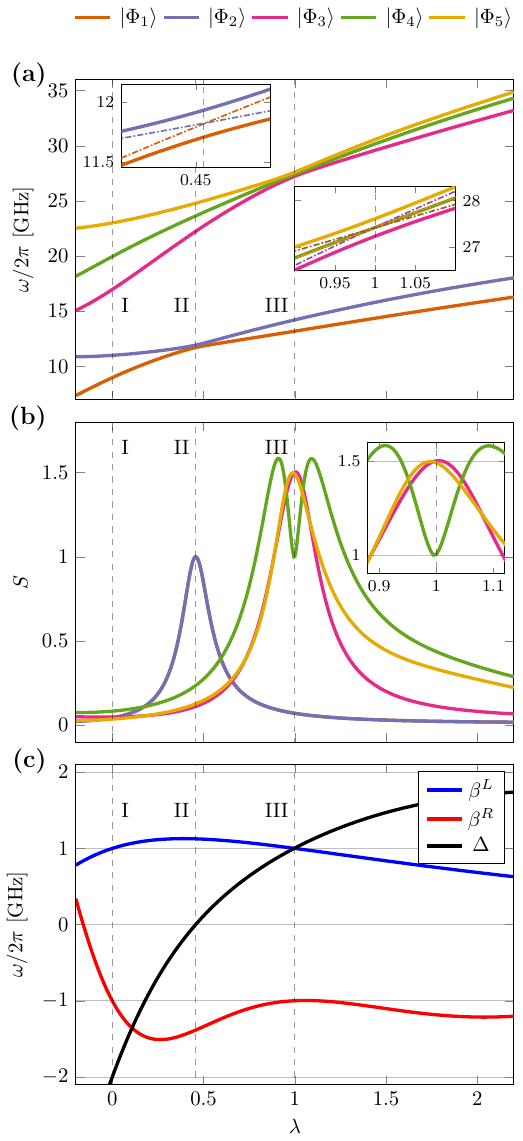}
    \caption{
        (a) Transition frequencies from the ground state of the five lowest excited energy eigenstates, as functions of the
        configurational parameter $\lambda$.
        Solid lines correspond to the transition energies of the
        full Hamiltonian. In the insets we have included thin dashed lines for the transition energies of the non-interacting Hartree product states.
        (b) Von Neumann entropies of the same five eigenstates as functions of the configurational parameter $\lambda$. The entropy is calculated with the binary (base-2) logarithm.
        (c) Anharmonicites of the left ($\beta^L$) and right ($\beta^R$)
        wells as functions of the configurational parameter $\lambda$, as
        well as the detuning $\Delta = \omega^R - \omega^L$ between
        the two wells.
        Marked in all subfigures are configurations I, II and III at
        their respective $\lambda$ values of $0$, $0.46$ and $1$.
    }
    \label{fig:crossings-entropies-anharmonicities}
\end{figure}

Figure~\ref{fig:crossings-entropies-anharmonicities} shows CI results
for the two-body energy spectrum for the five lowest excited states,
the corresponding entanglement entropies, as well as the
anharmonicities and detuning of the wells as a function of the
configurational parameter $\lambda$. Two avoided crossings are clearly
observed in the spectrum in
Figure~\ref{fig:crossings-entropies-anharmonicities}(a): a triple
avoided crossing at $\lambda=1$ between the three highest energy
states, and an avoided crossing between the two first excited states
at $\lambda \approx 0.46$, corresponding to configuration~II. In the
latter case we extract the coupling strength of $g_{\mathrm{CI}}
\approx 113$~MHz from the energy gap at the location of the avoided
crossing. This will be discussed in more details in
Sec.~\ref{sec:effective-model}.

Qualitatively, the impact of the Coulomb interaction on the system's
electrons can be understood in two steps. First, the electric field
created by one electron alters the potential energy experienced by the
other electron. This results in a modified effective potential trap,
which gives rise to the Hartree product states and their associated
energies. These non-interacting energies are depicted by the dashed
lines in the insets of
Fig.~\ref{fig:crossings-entropies-anharmonicities}(a). Second, in the
case of a voltage configuration which results in two or more Hartree
product states with the same energies, the residual Coulomb
interaction between the electrons lifts the degeneracy and leads to an
energy gap between the corresponding two-body energy eigenstates,
resulting in the above-mentioned avoided crossings. Far from the point
of degeneracy, the Hartree product states provide a good description
of the full two-body energy eigenstates. This can be observed, for
example, in configuration I at $\lambda=0$. In these configurations,
the calculated entropy values $S_n$ demonstrate minimal values,
indicating reduced correlations between the electrons. The entropy
values reach their maximum and align with theoretical values precisely
at the locations of the avoided crossings, as illustrated in
Fig.~\ref{fig:crossings-entropies-anharmonicities}(b).

A triple avoided crossing is observed in the higher energy states in
configuration III at $\lambda=1$, and arises due to the opposite signs
of the anharmonicities (see
Fig.~\ref{fig:crossings-entropies-anharmonicities}(c))~\cite{triple-avoided-crossing}. It
is worth mentioning that the anharmonicities vary across different
values of $\lambda$ since the linear interpolation of the voltages
does not guarantee that the properties of the system also behaves
linearly.

\begin{figure}
    \centering
    \includegraphics[width=\columnwidth]{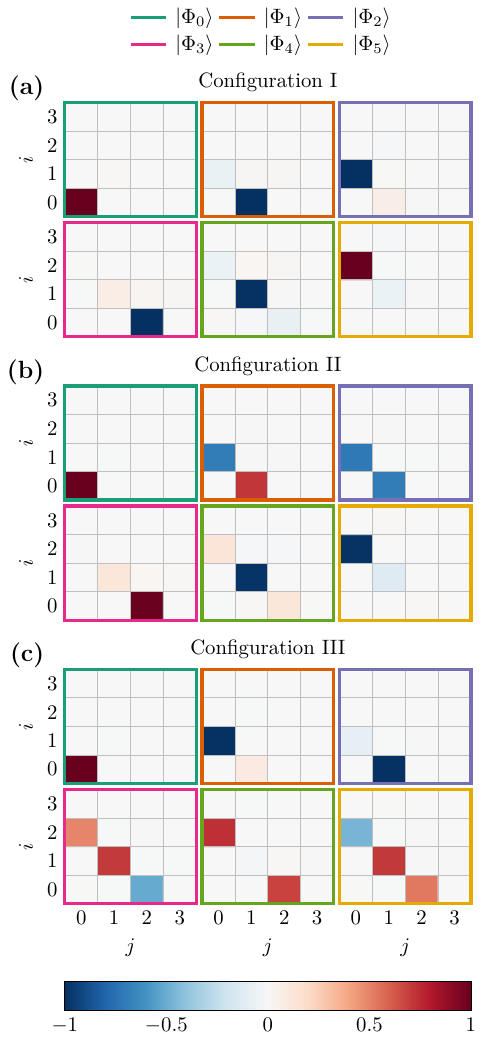}
    \caption{Two-body wave function coefficients $C_{ij,n}$ of the six lowest energy eigenstates for each configuration, as defined in Eq.~\eqref{eq:wave-function-ansatz}. The indices on the $x$ and $y$-axes correspond to the Hartree product states involved with each coefficient, so that the coefficient at tile ($i$,$j$) is multiplied with the product state $\ket{\phi^L_i\phi^R_j}$ and summed into the expansion of each energy eigenstate. (a) Coefficients for configuration I. Each energy eigenstate is well approximated by a single Hartree product state. (b) Coefficients for configuration II. The first and second excited eigenstates are close to maximally entangled. (c) Coefficients for configuration III. Here, the third, fourth and fifth excited eigenstates are entangled.}
    \label{fig:coefficients}
\end{figure}

The two-body coefficients $C_{ij,n}$ corresponding to the six lowest
energy eigenstates, as defined in the
ansatz~\eqref{eq:wave-function-ansatz}, are depicted in
Figure~\ref{fig:coefficients} for the three main configurations.
These coefficients demonstrate a good convergence of our optimization
algorithm towards the target wavefunctions presented in
Eqs.~\eqref{eq:target-states-II}~and~\eqref{eq:target-states-III}.  In
configuration I, the two-body eigenstates are effectively described by
single Hartree product states, indicating the suppression of
electron-electron correlations when the potential wells are detuned.
In contrast, the coefficients $C_{ij,n}$ for configurations II and III
reveal a high degree of entanglement, which is quantified using von
Neumann entropies.  A closer inspection of these coefficients reveals
the presence of small, undesired Hartree terms in the two-body
wavefunctions. For instance, for the first excited state in
configuration I, shown in Figure~\ref{fig:coefficients}(a), we find
\begin{equation}
    \ket{\Phi_1}_\mathrm{I} \approx \sqrt{0.995} \ket{01} + \sqrt{0.005} \ket{10}
\end{equation}
with a corresponding entropy of $S_1 \approx 0.04$. Furthermore, we
find a small mixing in states $\ket{\Phi_3}$, $\ket{\Phi_4}$, and
$\ket{\Phi_5}$, indicating residual correlations between the two
electrons through interactions with higher energy states. The degree
of these remaining correlations, quantified by the entropies $S_n$,
show small but non-zero values for all excited energy states. The
underlying factors contributing to these observations will be
discussed within the framework of the effective Hamiltonian model
presented in the following subsection.

For the first two excited states in configuration II, shown in
Fig.~\ref{fig:coefficients}(b), the many-body wavefunctions are
approximately described by
\begin{equation}
    \begin{aligned}
        \ket{\Phi_1}_\mathrm{II} &\approx \sqrt{0.513} \ket{01} - \sqrt{0.487} \ket{10},\\
        \ket{\Phi_2}_\mathrm{II} &\approx \sqrt{0.487} \ket{01} + \sqrt{0.513} \ket{10},
    \end{aligned}
\end{equation}
which are almost identical to the maximally entangled states in
Eq.~\eqref{eq:target-states-II}. The entropy for these entangled
states reach a maximum value of 1, as seen in
Fig.~\ref{fig:crossings-entropies-anharmonicities}(b). Here too, none
of the higher excited states can be entirely described by single
product states indicating to a presence of small residual
correlations. The entropies of the eigenstates $\ket{\Phi_3}$,
$\ket{\Phi_4}$ and $\ket{\Phi_5}$ for configuration II are around
$0.11$, $0.23$ and $0.13$, respectively.

We display the coefficients of the energy eigenstates for
configuration III in Figure~\ref{fig:coefficients}(c). The three
states involved in the triple avoided crossing are close to the target
states given in Eq.~\eqref{eq:target-states-III}:
\begin{align}
    \ket{\Phi_3}_\mathrm{III} &\approx - \sqrt{0.259} \ket{02} + \sqrt{0.244} \ket{20} + \sqrt{0.497} \ket{11},\nonumber\\
     \ket{\Phi_4}_\mathrm{III} &\approx \sqrt{0.461} \ket{02} + \sqrt{0.538} \ket{20} - \sqrt{0.001} \ket{11},\nonumber\\
    \ket{\Phi_5}_\mathrm{III} &\approx \sqrt{0.279} \ket{02} - \sqrt{0.218} \ket{20} + \sqrt{0.502} \ket{11}.
\end{align}
In this configuration, however, an unwanted coupling is present in the
first and second excited eigenstates $\ket{\Phi_1}$ and
$\ket{\Phi_2}$. The degree of entanglement for these states is rather
weak, as seen at $\lambda=1$ in
Fig.~\ref{fig:crossings-entropies-anharmonicities}(b); both
eigenstates have an entropy of around $0.07$.

We note that Fig.~\ref{fig:coefficients} demonstrates that the Hartree
states serve as approximate Schmidt states for all the energy
eigenstates of our system, as mentioned in
section~\ref{sec:entanglement}. This is clearly seen from this figure
since each row and column have approximately one non-zero coefficient.

For the sake of visualization, we include in
Appendices~\ref{app:Hartree-method}~and~\ref{app:particle-density} the
probability distributions of the one-body Hartree basis states
(Fig.~\ref{fig:Hartree-basis}) and the particle densities of the
two-body energy states (Fig.~\ref{fig:densities}) for each of the
three configurations.

\subsection{Effective Hamiltonian}\label{sec:effective-model}
    
In addition to the numerical results above, we present a simplified
model of the system to provide an intuitive understanding of the
underlying coupling mechanism between the two electrons. For this
purpose we expand both the electrostatic potential terms and the
Coulomb interaction in our model Hamiltonian
(Eq.~\eqref{eq:soft-coulomb}) around equilibrium positions $x_L$ and
$x_R$ for the two electrons. These equilibrium positions are defined
so that the first order terms in the displacements $\Delta{x}_i$
cancel each other, leaving only terms of second order and higher.

The Taylor expansion of the electrostatic potential around the
equilibrium positions results in harmonic traps $\omega_i^2
\Delta{x}_i^2/2$, with frequencies defined by the curvature of the
electrostatic potential at the equilibrium positions. The Coulomb
interaction between the two electrons can also be expanded in terms of
the displacements $\Delta{x}_i$. Considering only up to second-order
terms we obtain
    \begin{align}
        \frac{\kappa}{|x_1 - x_2|} \approx \frac{\kappa}{d}\Big ( 1 - \frac{\Delta x_1 - \Delta x_2}{d} + \frac{(\Delta x_1 - \Delta x_2)^2}{d^2} \Big ),
        \label{eq:coulomb-taylorseries}
    \end{align}
    where $d = x_R - x_L$ is the distance between the two electrons in equilibrium. The total potential energy of the system in displacement-dependent terms takes the form
    \begin{align}
        U_\mathrm{C} \approx \sum_{i=\mathrm{1,2}} \frac{(\omega_i^2 + \omega_\mathrm{C}^2)}{2}\Delta x_i^2 + \omega_\mathrm{C}^2\Delta x_1\Delta x_2,
       \label{eq:coulomb_displacement}
    \end{align}
where $\omega_\mathrm{C}^2=2\kappa/d^3$. The first term in this
equation describes how the Coulomb interaction effectively modifies
the potential wells from the electrostatic potential. This is similar
to the Hartree method since it computes an effective mean potential
for each electron, created by the other electron in the system,
however it is also different in that it treats the electrons as point
particles instead of quantum particles. The last term in
Eq.~(\ref{eq:coulomb_displacement}) gives rise to correlations between
the two electrons. By introducing canonical transformations for the
displacements and applying the rotating wave approximation, the
Hamiltonian of the system takes the form
    \begin{align}
        \hat{H} \approx \omega^L \hat{a}^{\dagger}\hat{a} + \omega^R \hat{b}^{\dagger}\hat{b} + g(\hat{a}^{\dagger}\hat{b} + \hat{a}\hat{b}^{\dagger}),
        \label{eq:coulomb_simpleH}
    \end{align}
where $\hat{a}$ and $\hat{b}$ are ladder operators of displacement in the left and right wells respectively,
$(\omega^L)^2 = \omega_1^2 + \omega_\mathrm{C}^2$ and $(\omega^R)^2 =
\omega_2^2 + \omega_\mathrm{C}^2$ are modified vibrational frequencies
and $g=\omega_\mathrm{C}^2/2\sqrt{\omega^L\omega^R}$ describes the
interaction strength.

This Hamiltonian is diagonalized by a standard Bogoliubov
transformation $U_\theta = \exp(\theta (\hat{a}^{\dagger}\hat{b} - \hat{a} \hat{b}^{\dagger}))$
with a rotation angle $2\theta = \arctan(2g/\Delta)$~\cite{blais2020circuit}. The resulting Hamiltonian takes the diagonal form $\hat{H} = \Omega^{+} \hat{\alpha}^{\dagger}\hat{\alpha} + \Omega^{-}\hat{\beta}^{\dagger}\hat{\beta}$. Here $\hat{\alpha}$ and $\hat{\beta}$ are the transformed ladder operators, and
the eigenfrequencies of the corresponding hybridized modes are given by
\begin{align}
        \Omega^{\pm} = \frac{1}{2} \Big (\omega^L + \omega^R \pm \sqrt{4 g^2 + \Delta^2} \Big ).
        \label{eq:coulomb_hybridmodes}
\end{align}
$\Delta = \omega^L - \omega^R$ is the detuning between the two wells as defined in Sec.~\ref{sec:config-search}.
    
Given the multilevel nature of electronic states in each well, one has
to carefully treat the unitary transformation of the effective
Hamiltonian in Eq.~(\ref{eq:coulomb_simpleH}). Including the
anharmonicity of each oscillator as additional terms $\beta^L\hat{a}^{\dagger}\hat{a} (\hat{a}^{\dagger}\hat{a} - 1)/2$ and $\beta^R \hat{b}^{\dagger}\hat{b} (\hat{b}^{\dagger}\hat{b} - 1)/2$ in the Hamiltonian, which corresponds to including quartic terms in the expansion of the electrostatic potential, results in correlations emerging from interactions between the higher energy states. After
performing a Bogoliubov transformation $U_{\theta}$ similar to the one
above, the term corresponding to the anharmonicities takes the form
\begin{align}
   \hat{H}_{ZZ} = \frac{\zeta}{2} \hat{\alpha}^{\dagger}\hat{\alpha} \hat{\beta}^{\dagger}\hat{\beta},
   \label{eq:zz-interaction}
\end{align}
where $\zeta$ is given by
\begin{align}
        \zeta = \sqrt{2} g \Big ( \tan{\frac{\theta_R}{2}} - \tan{\frac{\theta_L}{2}} \Big ),
        \label{eq:zz-coupling-2}
\end{align}
with $\tan{\theta_{L \atop R}} = 2 \sqrt{2} g / (\Delta \pm \beta^{L
  \atop R})$~\cite{triple-avoided-crossing}.  The quantity $\zeta$
corresponds to the energy shift defined in Eq.~\eqref{zz-coupling},
and is the result of interactions between the $\ket{20}$ and
$\ket{02}$ states and the $\ket{11}$ state.
    
\begin{figure}
        \centering
        \includegraphics[width=\columnwidth]{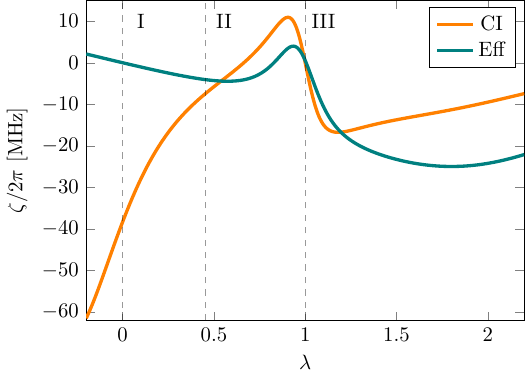}
        \caption{
            The energy difference $\zeta$ defined in Eq.~\eqref{zz-coupling}.
            The orange line represents results from the numerical solution of the full two-body problem, whereas the teal line shows $\zeta$ values calculated from the effective Hamiltonian approach. Frequencies obtained from Hartree energies and a fixed coupling strength determined at the avoided crossing of configuration II were used in the effective model calculations.
            Configurations I, II and III are marked with vertical dashed lines
            at their respective $\lambda$ values of $0$, $0.46$ and $1$.
        }
        \label{fig:zeta}
\end{figure}

In general, unwanted correlations from this type of interactions lead
to a conditional phase accumulation on the electron's states as
discussed in section~\ref{sec:gate-operation}. In Fig.~\ref{fig:zeta}
we show calculations of $\zeta$ from our CI calculations and from the
effective Hamiltonian approach. In the framework of the effective
Hamiltonian this quantity strongly depends on the relative signs of
the anharmonicities, which can be seen from the expression
\eqref{eq:zz-coupling-2}. For small $\theta_{L \atop R} \ll 1$ values,
which are realized in configuration I, the coupling strength can be
approximated by $\zeta \approx 2 g^2/(\Delta + \beta^L) - 2
g^2/(\Delta - \beta^R)$. This vanishes at equal but opposite sign
anharmonicities of two electrons. However, the CI calculations show a
strong deviation of $\zeta$ from the predictions based on the
effective model for well configurations $\lambda < 0.5$ and $\lambda >
1.5$ (see Fig.~\ref{fig:zeta}). We argue that these residual
correlations appear due to a complexity in the shape of the potential
wells. Nonlinearities on the localization length scale of electrons
requires us to include of higher-order terms in the Taylor expansion
of the electrostatic and Coulomb potentials. These terms, together
with the anharmonicities, can change for different voltage
configurations, a feature which further complicates our model. These
intricacies are inherent to the electrostatic field generated by the
array of electrodes in the micro-device considered. Potentially, the
$ZZ$ coupling strength can be included in the configurational search
as another minimization parameter to further suppress such
correlations between the two electrons.

The effective Hamiltonian presented in Eq.~(\ref{eq:coulomb_simpleH}),
represents one of the most elementary models for describing coupled
two-qubit systems.  This model finds widespread use in various
superconducting qubit architectures.  Its simplicity in form
facilitates the mapping of diverse entangling gates from these
platforms to our system.  However, alongside the simplicity of this
model, we have illustrated its limitations when comparing its
predictions with the results of full configuration interaction (CI)
calculations.  These limitations need to be handled thoughtfully in
order to account for all potential sources of entanglement.
Addressing these limitations becomes crucial for providing an accurate
and complete description of the entanglement dynamics within the
system and will be the scope of future work.

\section{Conclusions} 

The results presented in this work highlight how the Coulomb
interaction can induce motional entanglement between electronic states
localized in separate wells above the surface of superfluid helium. To
find the optimal specific device parameters for generating the
entangled states, we have developed an optimization method based on
many-body methods like full-configuration interaction (CI)
theory~\cite{CramerFCI} together with effective optimization
algorithms. Our optimization methodology allows us to determine the
optimal voltage configuration on the device electrodes needed to
generate entanglement. In this way, the many-body-physics-based
methodology we have developed has the potential to serve as a valuable
tool to guide experimental work and inform future device design.
        
As an illustration, in this work we examined three distinct device
parameter configurations (I, II, and III), leading to different types
of entanglement between the two electrons. The tunability of the
micro-device allows us to adjust the applied voltages and dynamically
create highly anharmonic electrostatic traps, even with varying signs
of anharmonicity. This tunability offers precise control over the
potential landscape experienced by the electrons and allows for the
tailoring of trapping potentials for specific experimental
requirements, such as the experimental realization of specific gates
and operations on the electronic qubits. Additionally we employed an
effective Hamiltonian to approximate the two-electron system, which
was in turn compared with our exact CI calculations, allowing us to
investigate the limitations of the approximations used to construct
this effective model. This comparison holds promise for a more
detailed understanding of errors in the simulation of quantum devices
based on this trapped electron system.
    
Finally, recent theoretical investigations have explored the dynamics
and decoherence of electron spins above the surface of liquid
helium~\cite{dykman2023spin}. These studies considered the role of
spin-orbit interactions, which can be artificially enhanced by
applying a spatially inhomogeneous magnetic field parallel to the
helium surface. In future studies the methodology developed in our
work can be extended to investigate entangling interactions between
spins, devices containing spatially varying magnetic fields, as well
as dynamical driving fields to investigate the time-dependence of
entangled charge states and spin states.
    
In addition to studies of the time evolution of these quantum
mechanical systems and thereby the temporal evolution of entangled
states, we plan to extend our studies to more than two particles, with
the aim to explore the experimental realization of many-body
entanglement for electrons above the surface of liquid helium and
solid neon. The hope is that these theoretical tools can guide studies
of entanglement, development of experimental devices and realization
of quantum gates and circuits for systems of many trapped and
interacting electrons.

\begin{acknowledgments}
We are grateful to M.I. Dykman and S.A. Lyon for illuminating
discussions. The work of MHJ is supported by the U.S. Department of
Energy, Office of Science, office of Nuclear Physics under grant
No. DE-SC0021152 and U.S. National Science Foundation Grants
No. PHY-1404159 and PHY-2013047. JP acknowledges support from the
National Science Foundation via grant number DMR-2003815 as well as
the valuable support of the Cowen Family Endowment at MSU. AKW
acknowledges support from the U.S. Department of Energy, Office of
Science, Basic Energy Sciences, Grant No. DE-SC0017889 and support
from MSU for a John A. Hannah Professorship. The work of NRB was
supported by a sponsored research grant from EeroQ Corp. JP and NRB
thank J.R. Lane and J.M. Kitzman for illuminating discussions. OL has
received funding from the European Union's Horizon 2020 research and
innovation program under the Marie Skłodowska-Curie grant agreement Nº
945371.
\end{acknowledgments}

\appendix

\section{Constructing the single-particle basis sets}\label{app:basis-set}
    
For our numerical calculations we use a pseudo-spectral basis, i.e., a
discrete variable representation (DVR), and adopt a linear
interpolation for the coupling constants $\alpha_i(x)$.  Specifically,
we use the one-dimensional $\sinc$-DVR basis suggested by
\citet{colbert-sinc}.  After dividing the Hamiltonian into two
distinguishable subsystems $L$ and $R$, as shown in
Eq.~\eqref{eq:one-body-hamiltonian}, we establish two $\sinc$-DVR
basis sets, one for each well.  We denote these basis functions by
$B^A = \qty{\chi^A_{\alpha}(x) \mid \alpha = 0, \dots, K^A}$ with the
corresponding quadrature of collocation points and weights $Q^A =
\qty{(x^A_{\alpha}, w^A_{\alpha}) \mid \alpha = 0, \dots, K^A}$ for $A
\in \qty{L, R}$.  The quadrature is uniform for the $\sinc$-DVR basis,
meaning that $w^A_{\alpha} = \Delta x^A$ and $x^A_{\alpha + 1} =
x^A_{\alpha} + \Delta x^A$ for all $\alpha$.

We let $x^L_{K^L + 1} = x_b = x^R_{0}$, i.e., the barrier is only
included as a quadrature point in the right system, and we let $\Delta
x = \Delta x^L = \Delta x^R$.  The $\sinc$-DVR functions are then
given by
        \begin{align*}
            \chi^A_{\alpha}(x) = \frac{1}{\sqrt{\Delta x}}
            \sinc\!\qty(
                \frac{x - x^A_{\alpha}}{\Delta x}
            ),
        \end{align*}
with
        \begin{align*}
            \sinc(x) = \begin{cases}
                \frac{\sin(\pi x)}{\pi x}, & x \neq 0, \\
                \hfil 1, & x = 0.
            \end{cases}
        \end{align*}
This means that $\chi^A_{\alpha} (x^A_{\beta}) = (\Delta x)^{-1/2}
\delta_{\alpha \beta}$ on the quadrature.  By restricting the grid on
each side only up to the barrier, we have effectively established an
infinite potential wall.  This means that the potentials given in
Eq.~\eqref{eq:potential-splitting} are altered to
        \begin{gather*}
            v^L(x) = \begin{cases}
                v(x), & x < x_b, \\
                \infty, & x \geq x_b,
            \end{cases}
            \\
            v^R(x) = \begin{cases}
                \infty, & x < x_b, \\
                v(x), & x \geq x_b.
            \end{cases}
        \end{gather*}
This forces each electron to remain in its own well, and might seem an
extreme limitation.  However, the results in our model are completely
unchanged, and it is much more computationally efficient and practical
to use two separate basis sets.

The matrix elements of the kinetic energy operator are given by
\cite{colbert-sinc}
        \begin{align*}
            t^A_{\alpha \beta}
            &= \mel*{\chi^A_{\alpha}}{-\frac{1}{2} \dv[2]{}{x}}{
                \chi^A_{\beta}
            }
            = \begin{cases}
                \frac{\pi^2}{6 (\Delta x)^2}, & \alpha = \beta, \\
                \frac{(-1)^{\alpha - \beta}}{
                    (\Delta x)^2 (\alpha - \beta)^2
                },
                & \alpha \neq \beta,
            \end{cases}
        \end{align*}
and the external potential is approximated using the quadrature rule, viz.,
        \begin{align*}
            v^A_{\alpha \beta}
            &= \mel*{\chi^A_{\alpha}}{\hat{v}^A(x)}{\chi^A_{\beta}}
            \\
            &\approx \Delta x
            \sum_{\gamma = 0}^{K^A}
            \chi^A_{\alpha}(x^A_{\gamma})
            v^A(x^A_{\gamma})
            \chi^A_{\beta}(x^A_{\gamma})
            = \delta_{\alpha \beta} v^A(x^A_{\beta}),
        \end{align*}
that is, the potential is diagonal.  The matrix elements of the full
one-body Hamiltonian can then be written
        \begin{align*}
            h^A_{\alpha \beta}
            &= t^A_{\alpha \beta}
            + \delta_{\alpha \beta}
            v^A_{\beta},
        \end{align*}
where we have defined the diagonal potential matrix elements
$v^A_{\beta} \equiv v^A(x^A_{\beta})$.

To evaluate the two-body Coulomb interaction, we examine the matrix
elements of tensor products of DVR-states., i.e.,
$\ket*{\chi^L_{\alpha} \chi^R_{\beta}} = \ket*{\chi^L_{\alpha}}
\otimes \ket*{\chi^R_{\beta}}$.  We also use the convention that
$\bra*{\chi^L_{\alpha} \chi^R_{\beta}} = \bra*{\chi^L_{\alpha}}
\otimes \bra*{\chi^R_{\beta}} = \ket*{\chi^L_{\alpha}
  \chi^R_{\beta}}^{\dagger}$ for the conjugate states.  We are able to
directly compute the matrix elements of the soft Coulomb interaction
operator using the quadrature rule.  The matrix elements are thus
        \begin{align*}
            u_{\alpha \beta, \gamma \delta}
            &= \mel*{\chi^{L}_{\alpha} \chi^{R}_{\beta}}{
                \hat{u}(x_1, x_2)
            }{\chi^L_{\gamma} \chi^R_{\delta}}
            \\
            &\approx \Delta x^L \Delta x^R
            \sum_{\sigma = 0}^{K^L} \sum_{\tau = 0}^{K^R}
            \chi^L_{\alpha}(x^L_{\sigma})
            \chi^R_{\beta}(x^R_{\tau})
            u(x^L_{\sigma}, x^R_{\tau})
            \nonumber
            \\
            &\qquad\times
            \chi^L_{\gamma}(x^L_{\sigma})
            \chi^R_{\delta}(x^R_{\tau})
            \\
            &= \delta_{\alpha \gamma} \delta_{\beta \delta}
            u(x^L_{\gamma}, x^R_{\delta}),
        \end{align*}
which is diagonal for each particle axis.  We label the matrix
elements of the diagonal Coulomb operator by $u^{LR}_{\gamma \delta}
\equiv u(x^L_{\gamma}, x^R_{\delta})$.

\section{The Hartree method}\label{app:Hartree-method}
    
In the Hartree method for two distinguishable particles we approximate
the ground state $\ket*{\Phi_0}$ of the full Hamiltonian $\hat{H}$ in
Eq.~\eqref{eq:hamiltonian} as the product state $\ket*{\Phi_0} \approx
\ket*{\Psi} = \ket*{\phi^L_0 \phi^R_0}$ under the constraint that the
Hartree orbitals are orthonormal, i.e., $\braket*{ \phi^A_0}{\phi^A_0}
= 1$.  This lets us set up the following Lagrangian
        \begin{align*}
            L
            = E_{H}
            - \lambda^L \qty(\braket*{\phi^L_0}{\phi^L_0} - 1)
            - \lambda^R \qty(\braket*{\phi^R_0}{\phi^R_0} - 1),
        \end{align*}
where $\lambda^A$ are Lagrange multipliers, and the Hartree energy
$E_H= \mel*{\Psi}{\hat{H}}{\Psi}$ is given by
        \begin{align*}
            E_H
            =
            \mel*{\phi^L_0}{\hat{h}^L}{\phi^L_0}
            + \mel*{\phi^R_0}{\hat{h}^R}{\phi^R_0}
            + \mel*{
                \phi^L_0 \phi^R_0
            }{u}{
                \phi^L_0 \phi^R_0
            }.
        \end{align*}
Our next objective is to minimize the Lagrangian with respect to the
Hartree states and the multipliers.  To do this we expand the Hartree
states as a linear combination of $\sinc$-DVR states:
        \begin{align}
            \ket*{\phi^A_i}
            = \sum_{\alpha = 0}^{K^A}
            B^A_{\alpha i} \ket*{\chi^A_{\alpha}},
            \label{eq:hartree-expansion}
        \end{align}
and minimize  with respect to the coefficients $B^A_{\alpha i}$.
Computing $\partial L/ \partial {B^A_{\alpha 0}}^{*} = 0$ gives two
coupled eigenvalue equations
        \begin{equation}
            \begin{gathered}
                \sum_{\beta = 0}^{K^L}
                \underbrace{
                    \qty(
                        h^L_{\alpha \beta}
                        + \delta_{\alpha \beta}
                        \sum_{\gamma = 0}^{K^R}
                        \abs{B^R_{\gamma 0}}^2
                        u^{LR}_{\beta \gamma}
                    )
                }_{\equiv f^L_{\alpha \beta}}
                B^L_{\beta 0}
                = \lambda^L B^L_{\alpha 0},
                \\
                \sum_{\beta = 0}^{K^R}
                \underbrace{
                    \qty(
                        h^R_{\alpha \beta}
                        + \delta_{\alpha \beta}
                        \sum_{\gamma = 0}^{K^L}
                        \abs{B^L_{\gamma 0}}^2
                        u^{LR}_{\gamma \beta}
                    )
                }_{\equiv f^R_{\alpha \beta}}
                B^R_{\beta 0}
                = \lambda^R B^R_{\alpha 0},
            \end{gathered}
            \label{eq:hartree-equations}
        \end{equation}
that needs to be solved iteratively until self-consistency has been
achieved.  The Hartree matrices $f^A_{\alpha \beta}$ are defined as
everything inside the parentheses in the equations above.  By
diagonalizing the Hartree matrices, we obtain $K^A$ eigenvalues and
eigenvectors, not just the lowest pair $\lambda^A$ and $B^A_{\alpha
  0}$.  We select the $N^A+1$ lowest eigenvectors, which gives us the
set $P^A=\qty{\ket{\phi^A_{i}}\mid i=0,\dots,N^A}$, where $N^A\ll
K^A$.

\begin{figure}[h!]
            \includegraphics[width=0.99\columnwidth]{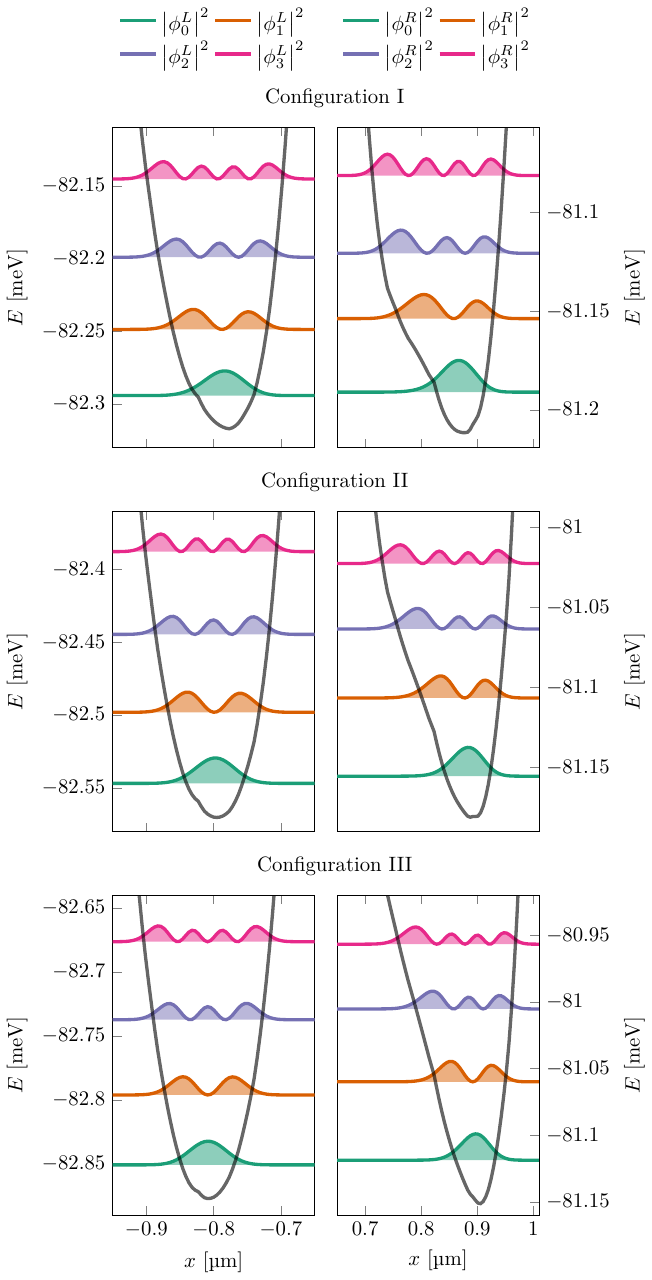}
            \caption{Probability distributions of the
            first four Hartree states in the left and right wells for all three configurations, shifted upwards with their Hartree transition frequencies from the respective ground states (in teal).
            The effective potentials for each subsystem, i.e., the well potential set up by the electrodes plus the mean-field contribution from the other electron,
            are plotted as dashed lines. As in Fig.~\ref{fig:densities} the left y-axis applies to the transition frequencies, while the right y-axis applies to the potentials, here for both panes.}
            \label{fig:Hartree-basis}
        \end{figure}

The equations we solve are $\hat{f}^A \ket*{\phi^A_i} = \epsilon^A_i
\ket*{\phi^A_i}$, where $\hat{f}^A$ is the Hartree matrix defined
earlier, and $\epsilon^A_i$ are the eigenvalues with the corresponding
eigenvectors $\ket*{\phi^A_i}$. These eigenvalues describe the energy
felt by a single particle trapped in one of the wells under the
influence of a charge in the other well. Formulated in terms of the
coefficients, the equations are
        \begin{align*}
            \sum_{\beta = 0}^{K^A}
            f^A_{\alpha \beta} B^A_{\beta i}
            = \epsilon^A_i B^A_{\alpha i},
        \end{align*}
with $f^A_{\alpha \beta}$ being the Hartree matrices from
Eqs.~\eqref{eq:hartree-equations}.  These equations are solved
iteratively until a convergence of $\abs{\epsilon^{A, (k + 1)}_i -
  \epsilon^{A, (k)}_i} < \delta \epsilon$ with $\delta \epsilon =
\num{1e-10}$ has been reached.  Here $k$ corresponds to an iteration
number.  We choose $f^{A, (0)}_{\alpha \beta} = h^A_{\alpha \beta}$ as
an initial state such that
        \begin{align*}
            \sum_{\beta = 0}^{K^A}
            h^A_{\alpha \beta} B^{A, (0)}_{\beta i}
            = \epsilon^{A, (0)}_i B^{A, (0)}_{\alpha i}.
        \end{align*}
For the three target configurations found through numerical
optimization in this work, the probability distributions of the first
four Hartree states in each well are plotted in
Fig.~\ref{fig:Hartree-basis}.

\section{Full configuration-interaction for two distinguishable particles}\label{app:fci}
    
Once the Hartree equations are solved, we obtain the coefficients
$B^A_{\alpha i}$, which allow us to construct the Hartree basis $P^A$
from the $\sinc$-DVR basis $B^A$ using
Eq.~\eqref{eq:hartree-expansion}.  We can perform a basis
transformation from the $\sinc$-DVR basis to the smaller Hartree basis
by using relations:
        \begin{gather}
            h^A_{ij}
            = \sum_{\alpha = 0}^{K^A} \sum_{\beta = 0}^{K^A}
            B^{A*}_{\alpha i} B^A_{\beta j} h^A_{\alpha \beta},
            \label{eq:hartree-ob-mels}
            \\
            u_{ij, kl}
            = \sum_{\alpha = 0}^{K^L} \sum_{\beta = 0}^{K^R}
            B^{L*}_{\alpha i}
            B^{R*}_{\beta j}
            B^{L}_{\alpha k}
            B^{R}_{\beta l}
            u^{LR}_{\alpha \beta},
            \label{eq:hartree-tb-mels}
        \end{gather}
where greek letters denote matrix elements in the $\sinc$-DVR basis,
and latin letters are for the Hartree basis.

Upon inserting the wave function ansatz into the time-independent
Schrödinger equation and projecting onto a two-body state
$\bra*{\phi^L_i \phi^R_j}$, we get:
        \begin{align*}
            \mel*{\phi^L_i \phi^R_j}{\hat{H}}{\Phi_n}
            &=
            \sum_{k = 0}^{N^L} \sum_{l = 0}^{N^R}
            H_{ij, kl} C_{kl, n}
            = C_{ij, n} E_n,
        \end{align*}
where $H_{ij, kl} \equiv \mel*{\phi^L_i \phi^R_j}{\hat{H}}{\phi^L_k
  \phi^R_l}$ are the matrix elements of the Hamiltonian in the Hartree
product basis.  The solution of this eigenvalue equation yields the
coefficients $C_{ij,n}$, where each column corresponds to an
eigenstate $\ket{\Phi_n}$ with corresponding eigenenergy $E_n$. The
matrix elements of the two-body Hamiltonian can be expressed as:
        \begin{align*}
            H_{ij,kl}
            &= h^L_{ik} \delta_{jl}
            + \delta_{ik} h^R_{jl}
            + u_{ij, kl},
        \end{align*}
where the one- and two-body matrix elements in the Hartree basis are
shown in
Eqs.~\eqref{eq:hartree-ob-mels}~and~\eqref{eq:hartree-tb-mels}.

\section{The von Neumann entropy}\label{app:von-neumann-entropy}
    
The von Neumann entropy is defined by
        \begin{align*}
            S
            = -\tr\!\qty(
                \hat{\rho}
                \log_2(\hat{\rho})
            ),
        \end{align*}
where $\hat{\rho}$ is the density operator.  The entropy of the
eigenstates $\ket*{\Phi_n}$ will be zero as they are pure states.
However, the entropy of the reduced subsystems ($L$ and $R$) of
$\ket*{\Phi_n}$ will in general not be zero.  Each subsystem will have
the same entropy, and any non-zero entropy can be attributed to
entanglement.  We can evaluate the entanglement entropy by bypassing
the construction of the reduced density operator and using the Schmidt
decomposition instead.  Specifically, for a given two-body wave
function $\ket*{\Psi}$ expressed in terms of the Hartree product
states,
        \begin{equation*}
            \ket*{\Psi} = \sum_{k = 0}^{N^L} \sum_{l = 0}^{N^R} C_{kl} \ket*{\phi^L_k \phi^R_l},
        \end{equation*}
we can perform a singular value decomposition of the two-body
coefficients, $C_{kl} = \sum_{p = 0}^{\tilde{N}} U_{kp} \sigma_{p}
V^{*}_{lp}$, to obtain
        \begin{equation*}
            \ket*{\Psi} = \sum_{p = 0}^{\tilde{N}} \sigma_{p} \ket*{\psi^L_p \psi^R_p},
        \end{equation*}
        where
        \begin{gather*}
            \ket*{\psi^L_p}
            \equiv \sum_{k = 0}^{N^L}
            U_{kp} \ket*{\phi^L_k},
            \qquad
            \ket*{\psi^R_p}
            \equiv \sum_{l = 0}^{N^R}
            V^{*}_{lp} \ket*{\phi^R_l},
        \end{gather*}

are the Schmidt states, $\tilde{N}$ is either $N^L$ or $N^R$ depending
on the definition of the singular value decomposition, and $\sigma_p$
are the singular values with $\sigma_p^2$ representing the occupation
of the single-particle states $\ket*{\psi^L_p}$ and $\ket*{\psi^R_p}$.
Using the singular values, we can then compute the von Neumann entropy
of $\ket{\Psi}$ as follows:
        \begin{align*}
            S = -\sum_{p = 0}^{\tilde{N}}
            \sigma_p^2 \log_2(\sigma_p^2).
        \end{align*}

\section{Particle densities}\label{app:particle-density}
\begin{figure}[h!]
            \centering
            \includegraphics[width=0.99\columnwidth]{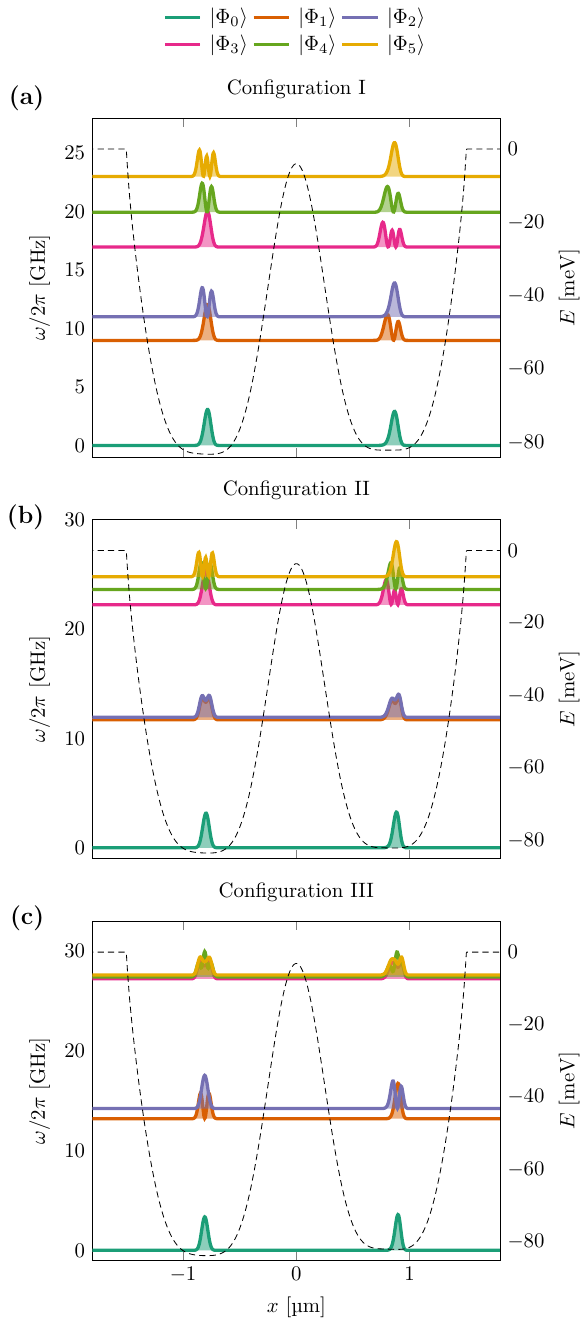}
            \caption{Calculated particle densities of the first six eigenstates of the full two-body Hamiltonian for configurations I (a), II (b) and III (c). Each state is shifted upwards with its transition frequency from the ground state (in teal). The electrostatic potential wells are shown by dashed lines.
            Note that these are hybrid plots; the left y-axis gives the unit scale for the transition frequencies, while the right y-axis gives the unit scale for the electrostatic potential energy.}
            \label{fig:densities}
        \end{figure}

For the state $\ket{\Psi}$ above, we can compute the particle density
by
        \begin{align*}
            \rho(x)
            &= \int \dd{y} \abs{\Psi(x, y)}^2
            + \int \dd{y} \abs{\Psi(y, x)}^2
            \\
            &= \sum_{i, j = 0}^{N^L}
            \sum_{l = 0}^{N^R} C^{*}_{il} C_{jl}
            {\phi^{L}_{i}}^{*}(x) \phi^L_{j}(x)
            \nonumber \\
            &\qquad
            + \sum_{i, j = 0}^{N^R}
            \sum_{k = 0}^{N^L} C^{*}_{ki} C_{kj}
            {\phi^{R}_{i}}^{*}(x) \phi^R_{j}(x),
        \end{align*}
which collapses to the electron density in the case of
indistinguishable particles. The calculated particle densities for the
three target configurations found through numerical optimization in
this work are shown in Fig.~\ref{fig:densities}.
        
\section{Finding optimal well configurations}\label{app:opt-well-conf}
        
In this appendix, we present a way of finding the optimal
configuration for single motional qubit rotations, configuration I, as
well as the optimal configurations for two-qubit operations,
configuration II and III.  These are found by expressing our
configurational search in terms of an optimization problem. The seven
voltages of the potential from Eq.~\eqref{eq:trap}, denoted
$\boldsymbol{V}$, will be varied to find the optimal solution.  We
note that due to the flexibility provided by the potential, the
optimization landscape consists of several local minima and the
resulting voltages are therefore somewhat arbitrary.  The same can
also be said for the path between configurations.  We have chosen
configurations and a path such that our results resemble those of
Fig.~2(b) by \citet{triple-avoided-crossing}, but we stress that our
model allows for vastly different solutions.  We select a fixed
anharmonicity with equal magnitude and opposite sign for each well,
and try to tune the wells such that only the detuning between each
well is altered.

Configuration I is a configuration for which the transition
frequencies are distinct, but at the same time within the working
range of 5--\SI{15}{\giga\hertz} of the read-out resonators.
Furthermore, we aim for the anharmonicities in the left and right
well, denoted as $\beta^L / 2\pi$ and $\beta^R / 2\pi$, to have equal
magnitudes but with opposite signs. This adjustment is made to
eliminate $ZZ$ crosstalk and facilitate a high on-off ratio for the
implementation of controlled phase-gates. \cite{zz-suppression,triple-avoided-crossing}.
There are several possible candidates for
the transition frequencies and anharmonicities which satisfy these
requirements, and the candidates we end up with are a result of
performing the optimization process for a range of possible
candidates.  For the left well, we targeted a transition frequency
between the two lowest energy levels of $\omega^L / 2 \pi =
\epsilon^L_1 - \epsilon^L_0 = \SI{11}{\giga \hertz}$, and a
corresponding transition frequency for the right well of $\omega^R / 2
\pi = \epsilon^R_1 - \epsilon^R_0 = \SI{9}{\giga \hertz}$.  Here
$\epsilon^A_i$ are the Hartree eigenvalues, i.e., the single-particle
Hartree energies.  At the same time, we targeted anharmonicites of
$\beta^L / 2\pi = - \beta^R / 2 \pi = \SI{1}{\giga\hertz}$.  If we
were allowed to vary the transition frequencies and anharmonicities
independently and freely, a cost function with minima that coincide
with these properties is
        \begin{align}
            \label{eq:LossConfigurationI}
            C_\mathrm{I}(\boldsymbol{V}) &= (\omega^L (\boldsymbol{V}) / 2 \pi - \SI{11}{\giga \hertz} )^2 \notag \\
            &+ ( \omega^R (\boldsymbol{V}) / 2 \pi - \SI{9}{\giga \hertz} ) ^2 \notag \\
            &+ ( \beta^L(\boldsymbol{V}) / 2 \pi - \SI{1}{\giga \hertz} ) ^2 \notag \\
            &+ ( \beta^R(\boldsymbol{V}) / 2\pi + \SI{1}{\giga \hertz}) ^2,
        \end{align}
where $\omega^{A} (\boldsymbol{V}) / 2 \pi$ is the transition
frequency and $\beta^{A} (\boldsymbol{V}) / 2\pi$ is the anharmonicity
of the wells (with $A\in\{L,R\}$). To minimize
$C_\mathrm{I}(\boldsymbol{V})$ we evaluated its gradient with respect
to the voltages, that is,
$\nabla_{\boldsymbol{V}}C_\mathrm{I}(\boldsymbol{V})$, using the
Tensorflow machine learning library
\cite{tensorflow2015-whitepaper}. We then used a variation of the
gradient descent method with an adaptive learning rate based on the
ADAM algorithm~\cite{adamoptimizer}, to update the voltages. The
learning rate for the Adam optimizer was initially set to $10^{-4}$.

For configuration III, we want to tune into a triple degeneracy point
between the states $\ket{\Phi_3}$, $\ket{\Phi_4}$ and $\ket{\Phi_5}$.
This allows for the realization of a controlled-phase
gate~\cite{zz-suppression,zz-suppression2}.  In such a configuration,
we construct a cost function based on targeting the von Neumann
entropies of the eigenstates $\ket{\Phi_3}$, $\ket{\Phi_4}$ and
$\ket{\Phi_5}$ to be $S_3 = 1.5$, $S_4 = 1.0$ and $S_5 = 1.5$,
respectively, while the entropies of the lower eigenstates should be
kept minimal.  We targeted the same anharmonicities as for
configuration I, that is, $\beta^L/2\pi = -\beta^R/2\pi =
\SI{1}{\giga\hertz}$.  In order to end up with a configuration close
to configuration I in parameter space, we utilized the parameters for
configuration I, denoted $\boldsymbol{V}_\mathrm{I}$, as the initial
guess in the optimization algorithm.  Finally, we ensure that a linear
sweep of voltages from configuration I to configuration III passes
through configuration II by targeting a detuning
$\Delta_\mathrm{III}/2\pi = (\omega^L - \omega^R)/2\pi = \SI{-1}{\giga
  \hertz}$ for configuration III, as explained in section
\ref{sec:config-search}.  The cost function we will apply is given by
        \begin{align}
            \label{eq:LossConfigurationIII}
            C_\mathrm{III}(\boldsymbol{V}) &=
            S_1(\boldsymbol{V})^2 \notag \\
            &+ S_2(\boldsymbol{V})^2 \notag \\
            &+ (S_3(\boldsymbol{V}) - 1.5)^2 \notag \\
            &+ (S_4(\boldsymbol{V}) - 1.0)^2 \notag \\
            &+ (S_5(\boldsymbol{V}) - 1.5)^2 \notag \\
            &+ (\beta^L (\boldsymbol{V}) / 2\pi - \SI{1}{\giga \hertz} )^2 \notag \\
            &+ (\beta^R (\boldsymbol{V}) / 2\pi + \SI{1}{\giga \hertz} )^2 \notag \\
            &+ (\omega^L (\boldsymbol{V}) / 2 \pi - \omega^R (\boldsymbol{V}) / 2 \pi + \SI{1}{\giga \hertz})^2.
        \end{align}
We used the same optimization method and learning rate as for configuration I.

\section{Optimized electrode voltages}\label{app:voltages}
        
The explicit values of the electrode voltages obtained for the three
optimized configurations are shown in table~\ref{tab:voltages} below.
        
        \begin{table}[h!]
        \caption{Electrode voltages for the three configurations I, II, and III, resulting in a total electrostatic potential given by Eq.~\eqref{eq:trap}.}
        \begin{ruledtabular}
            \begin{tabular}{lrrr}
                & \text{I}~[\si{\milli\V}] & \text{II}~[\si{\milli\V}]
                & \text{III}~[\si{\milli\V}] \\
                $V_1$ & 297.45 & 296.41 & 295.17
                \\
                $V_2$ & 152.33 & 155.01 & 158.23
                \\
                $V_3$ & 344.36 & 342.63 & 340.56
                \\
                $V_4$ & -353.49 & -354.66 & -356.05
                \\
                $V_5$ & 345.93 & 345.26 & 344.46
                \\
                $V_6$ & 143.94 & 143.78 & 143.60
                \\
                $V_7$ & 302.24 & 302.81 & 303.49
                \\
            \end{tabular}
        \end{ruledtabular}
        \label{tab:voltages}
    \end{table}

\bibliography{paper}
\end{document}